\documentclass[pra,twocolumn,aps,superscriptaddress]{revtex4-1}
\usepackage{amsfonts}
\usepackage{mathtools}
\usepackage[svgnames,table]{xcolor}
\usepackage[T1]{fontenc}
\usepackage{txfonts}
\usepackage{graphicx}
\usepackage{lineno}

\usepackage{hyperref}
\hypersetup{
        colorlinks=true,
        citecolor=SteelBlue,
        filecolor=LimeGreen,
        linkcolor=SlateBlue,
        urlcolor=MediumPurple
}

\usepackage{dcolumn}
\usepackage{bm}
\usepackage{amsmath}
\usepackage{breqn}
\usepackage[bbgreekl]{mathbbol}
\usepackage{bbm}
\usepackage{mathrsfs}
\usepackage[permil]{overpic}

\newcommand{\comment}[2]{\textcolor{orange}{}}

\makeatletter
\let\cat@comma@active\@empty
\makeatother

\begin{document}
\preprint{APS/123-QED}
\title{Design of a tabletop interferometer with quantum amplification}
\author{Jiri Smetana}
\email{gsmetana@star.sr.bham.ac.uk}
\affiliation{Institute for Gravitational Wave Astronomy, School of Physics and Astronomy, University of Birmingham, Birmingham B15 2TT, United Kingdom}\author{Artemiy Dmitriev}
\email{admitriev@star.sr.bham.ac.uk}
\affiliation{Institute for Gravitational Wave Astronomy, School of Physics and Astronomy, University of Birmingham, Birmingham B15 2TT, United Kingdom}
\author{Chunnong Zhao}
\affiliation{OzGrav,University of Western Australia, 35 Stirling Highway, Crawley, WA 6009, Australia}
\author{Haixing Miao}
\email{haixing@tsinghua.edu.cn}
\affiliation{State Key Laboratory of Low Dimensional Quantum Physics, Department of Physics, Tsinghua University, Beijing, China}
\affiliation{Frontier Science Center for Quantum Information, Beijing, China}
\author{Denis Martynov}
\email{d.martynov@bham.ac.uk}
\affiliation{Institute for Gravitational Wave Astronomy, School of Physics and Astronomy, University of Birmingham, Birmingham B15 2TT, United Kingdom}
\date{\today}

\begin{abstract}
The sensitivity of laser interferometers is fundamentally limited by the quantum nature of light. 
Recent theoretical studies have opened a new avenue to enhance 
their quantum-limited sensitivity by using active parity-time-symmetric and phase-insensitive quantum amplification. 
These systems can enhance the signal response without introducing excess noise in the ideal case. However, such active systems must be causal, stable, and carefully tuned to be practical and applicable to precision measurements. In this paper, we show that phase-insensitive amplification in laser interferometers can be implemented in a tabletop experiment. The layout consists of two coupled cavities and an active medium comprised of a silicon nitride membrane and an auxiliary pump field. Our design relies on existing membrane and cryogenic technology and can demonstrate three distinct features: (i) the self-stabilized dynamics of the optical system, (ii) quantum enhancement of its sensitivity in the presence of the amplifier, and (iii) optical control of the amplifier gain. These features are needed to enhance the sensitivity of future interferometric gravitational-wave and axion detectors.
\end{abstract}

\maketitle

\section{\label{sec:intro}Introduction}
Improvements in the performance of gravitational-wave (GW) detectors continue to stretch the known boundaries of precision measurement. Ever since the first discovery of gravitational waves~\cite{BBHDetection}, there has been a concerted effort to enhance both the sensitivity and bandwidth of these detectors. These allow us to capture a wider range of astrophysical phenomena whose detection is only possible due to their GW emission, such as merger events between black holes~\cite{BBHDetection}, neutron stars~\cite{BNSDetection}, or both~\cite{BHNSDetection}. The current pinnacle of sensitivity is achieved by the Advanced LIGO~\cite{AdvLIGO} and Advanced Virgo~\cite{AdvVirgo} detectors and is limited over much of the spectrum by fluctuations brought about by the quantum nature of light~\cite{IntroQN}. Improvements beyond previous quantum-induced limitations in interferometric systems have been implemented already, ranging from changes to detector configuration (such as the introduction of signal recycling~\cite{SignalRecycling1998, SignalRecycling2001}) to implementing direct quantum-noise suppression techniques (such as the squeezed states of light~\cite{Squeezing1981,Squeezing2013,Squeezing2017,Squeezing2020a,Squeezing2020b}). However, there are reasons for further enhancements in the sensitivity and bandwidth of GW detectors. Continuous improvements in detector sensitivity will provide us with deeper and better localization~\cite{Local2011, Local2018}. Existing performance improvements have led to a faster growing catalog of GW sources~\cite{Catalogue1, Catalogue2, Catalogue3}, which allows us to obtain population statistics~\cite{Population}. Further increases in detector bandwidth can lead to the observation of high-frequency phenomena, such as remnant collapse, aloowing us to probe neutron star physics~\cite{NeutronStar2018, NeutronStar2019, Collapse}, and core collapse supernovae~\cite{CoreCollapse2009, CoreCollapse2013, CoreCollapse2019}.

There is ample motivation for increasing the sensitivity and bandwidth without sacrificing either or, ideally, improving both. The limits on these two properties are imposed by the quantum fluctuations of the light field itself~\cite{CramerRao2011, CramerRao2017}, combined with the response of the detector's optical cavities. It is difficult to achieve simultaneous improvements in both due to the constraints imposed by the Mizuno limit~\cite{Mizuno}, which shows an inverse relationship between the peak sensitivity and bandwidth of the optical system. One of the key insights into this limit is the generation of positive dispersion by the optical cavities present in the system. Several proposals have been made for enhancing the detector performance beyond the standard quantum-imposed constraint using an optomechanical filter cavity~\cite{Optomech2014,OptomechExternal2015,Optomech2019,OptomechConvert2019,OptomechSRC2021,Xiang_2020,OptomechTD2021}. This system has been variously analysed as a bandwidth-broadening device~\cite{OptomechExternal2015}, a white light cavity~\cite{OptomechTD2021} and a phase-insensitive filter~\cite{OptomechIdeal2021}. Proposals consider the implementation of the filter as an auxiliary cavity attached to existing detectors~\cite{OptomechExternal2015}, or as a conversion of the existing signal-recycling cavity~\cite{OptomechConvert2019}. A mathematically analogous system has also been proposed that consists of a purely optical implementation~\cite{OptomechOptical2021}. Many of these proposals consider an unstable system, requiring further active stabilization. In recent studies~\cite{Xiang_2020,OptomechTD2021,OptomechIdeal2021}, it has been argued that alternate configurations of the filter cavity and the signal read-out scheme can result in a stable system, which still retains sensitivity enhancement beyond the Mizuno limit.

We propose the tabletop layout that can verify the validity of quantum amplification models~\cite{OptomechConvert2019, OptomechOptical2021, OptomechTD2021}. In a scaled-down system analogous to the analysis in Ref.~\cite{OptomechTD2021}, which was applied to a contemporary GW detector, we make use of a coupled-cavity scheme that augments one cavity with a phase-insensitive amplifier. The amplifier performs a transformation of its input field $a$ according to the equation~\cite{Caves1982}
\begin{equation}\label{eq:quamp}
    b = G a + K n,
\end{equation}
where $b$ is its output mode, $n$ is the filter noise, $G$ is the amplifier gain~\footnote{We follow the formalism introduced by Caves in Ref.~\cite{Caves1982}. In Eq.~\ref{eq:quamp}, quantities $a$ and $b$ represent the input and output field and $n$ represents the added noise. They can be treated either as the annihilation operators of the corresponding modes of the electromagnetic field in the fully quantum picture or as dimensionless complex amplitudes in the semi-classical approach. We refer to all systems described by this equation as ``phase-insensitive amplifiers'', even though the ``gain'' magnitude $|G|$ can be smaller than or equal to one, i.e. the system can act as an attenuator or a phase shifter.}, and $K$ is the noise coupling coefficient related to $G$ according to the equation $|K|^2 = |G|^2 - 1$ in order to make the transformation unitary.

The effect relies on the ratio between the optomechanical coupling rate (between the filter cavity and the mechanical resonator) and the optical coupling rate (between the two cavities) being close to unity. Since the latter increases as the main cavity length decreases~\cite{NeutronStar2019}, a straightforward down-scaling of the kilometer-size design analyzed in~\cite{OptomechTD2021} to a tabletop experiment is not possible. Such an experiment, however, is essential for developing deeper understanding of the fundamental physics underlying the parity-time-symmetric quantum filtering before it can be applied to GW detectors. Other technical challenges of the tabletop configuration include accounting for the thermal noise introduced by the mechanical resonator, stabilizing the resonant frequency (locking) of the coupled cavity system, and providing effective mode matching between the small beam waist size for the optomechanical interaction and larger beam size required for the stability of a meter-scale setup.

We show how the challenges listed above can be overcome in a tabletop setup with an appropriate choice of parameters. The proposed interferometer implements an optomechanical interaction of the signal field with a Si$_3$N$_4$ membrane, which can achieve high mechanical quality factors of up to $10^9$~\cite{Rossi2018} at cryogenic temperatures (10\,K). The main goals of the proposed experiment are to (i) demonstrate the stability of the optical system with the quantum filter, (ii) measure the propagation of the signal and noise fields in the system (iii) prove that phase-insensitive filtering can improve the sensitivity of quantum-limited interferometric detectors. We outline the theory of quantum amplification in optical interferometers in Section~\ref{sec:theory} and find the optomechanical parameters suitable for tabletop demonstration in Section~\ref{sec:par}. We discuss the quantum performance of the setup in Section~\ref{sec:quantum}.

\section{\label{sec:theory}Dynamics of the proposed system}

\begin{figure*}[t]
\centering
\includegraphics[width=0.98\textwidth]{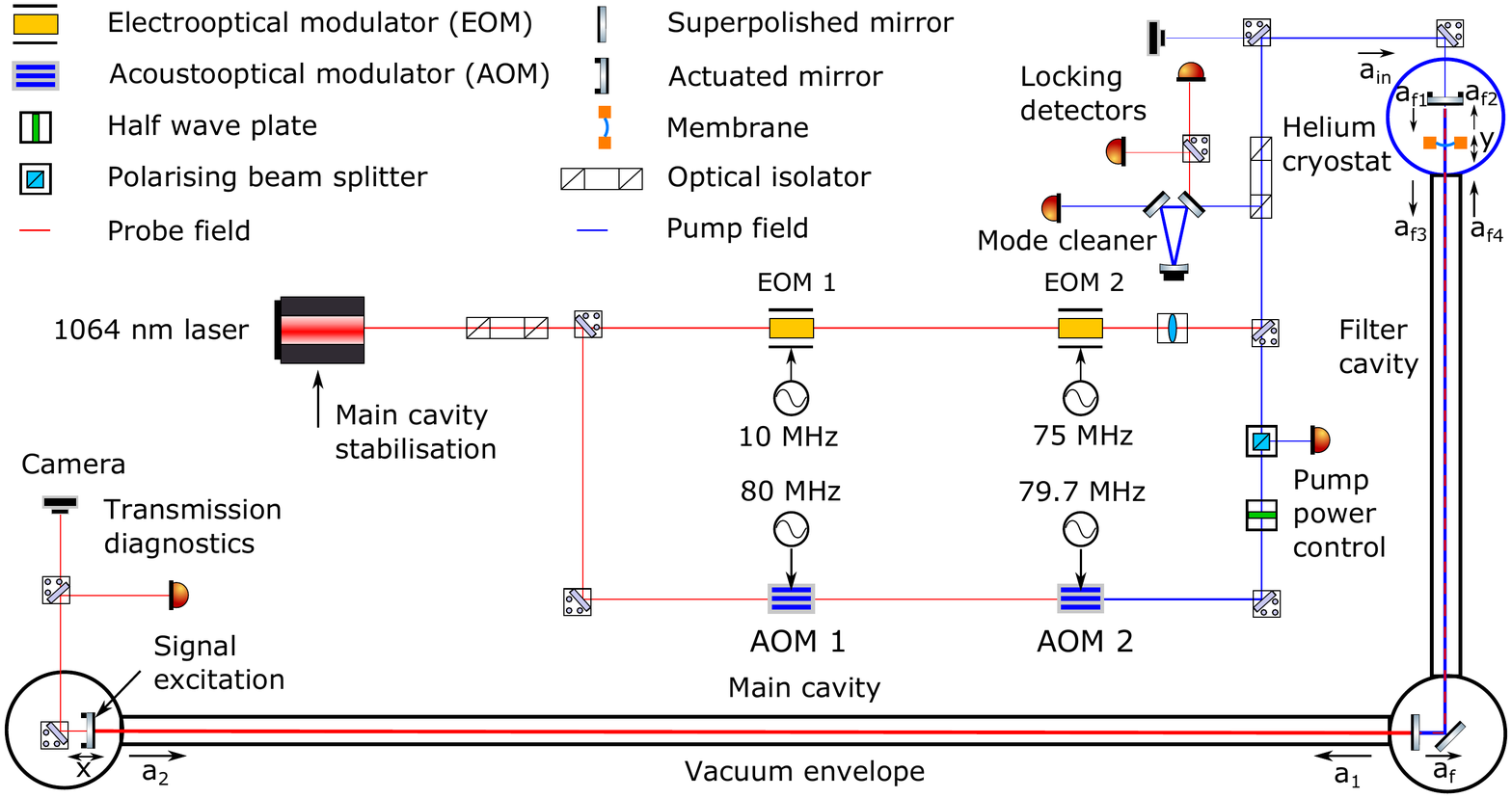}
\caption{Layout of the proposed experiment. It consists of the two coupled main and filter cavities. The setup includes control of the pump power for the filter cavity. The silicon nitride membrade embedded in the filter cavity implements the quantum amplification. The pump field is derived from the main laser with two acousto-optic modulators (AOMs). The driving frequency of AOM 2 is offset from the one of AOM 1 by the required offset between the pump and probe beams (303\,kHz for our experiment).
}
\label{fig:exp}
\end{figure*}

Our design consists of a coupled-cavity interferometer with a resonant mode, $\omega_0$, as shown in Fig.~\ref{fig:exp}. The optimised experimental parameters are listed in Table~\ref{table:par}. The signal field at frequencies $\omega_0 \pm \omega_s$ is produced inside the high-finesse main cavity and is further amplified inside the filter cavity. The amplification is achieved by a membrane with a mechanical mode at $\omega_m$ and an auxiliary pump field at frequency $\omega_0 + \omega_p = \omega_0 + \omega_m + \omega_{\rm OS}$, where $\omega_{\rm OS}$ is the frequency shift of the mechanical oscillator due to an optical spring in the filter cavity\,\cite{OptomechTD2021}.


\begin{table}[h]
\begin{center}
\caption{\label{table:par}Experimental parameters.}
\begin{ruledtabular}
\begin{tabular}{lcc}
 Parameter & Symbol	& Value \\
 \hline
 Main cavity length & $L_0$ & 4.1\,m \\
 Main cavity input coupler transmissivity & ${\cal T}_0$ & 30\,ppm \\
 Main cavity loss & $\epsilon_0$ & 10\,ppm \\
 Filter cavity length &	$L_f$ &	2\,m \\
 Filter cavity bandwidth &	$\gamma_f/2\pi$ & 30\,kHz \\
 Filter cavity input coupler transmissivity & ${\cal T}_f$ & 0.5\,\% \\
 Filter cavity loss & $\epsilon_f$ & 2000\,ppm \\
 Membrane eigenmode & ${\omega_m}/{2\pi}$ & 300\,kHz \\
 Motional mass & M & 40\,ng \\
 Membrane thickness & h & 50\,nm \\
 Membrane transmissivity & ${\cal T}_m$ & 0.8 \\
 Membrane temperature & $T$ & 10\,K \\
 Input pump power & $P_{\rm in}$ & 70\,mW \\
 Filter cavity power & $P_{\rm f}$ & 3.4\,W \\
 Pump frequency offset & $\omega_p / 2\pi$ & 303\,kHz
\end{tabular}
\end{ruledtabular}
\end{center}
\end{table}

Our layout is similar to a contemporary GW detector with the auxiliary signal recycling cavity tuned to broaden the antenna response at the expense of the gain at DC: the carrier field at the frequency $\omega_0$ is resonant in the arm cavity but anti-resonant in the signal recycling cavity. Our main cavity and filter cavity can be identified with the arm cavity and signal recycling cavity of the canonical GW detector, respectively.  The distinguishing feature of our layout is the silicon nitride (Si$_3$N$_4$) membrane embedded in the filter cavity. Among a vast diversity of optomechanical oscillators~\cite{Aspelmeyer_2014}, we choose the membrane because it can support relatively large beam sizes ($\sim 1$\,mm) and can exhibit high mechanical frequencies ($\sim 300$\,kHz) with a sufficiently high intrinsic tension. These properties make the technology readily applicable to the km-scale Advanced LIGO detectors without changing the g-factors of their signal recycling cavities.

In \cite{OptomechTD2021}, the mechanical resonator was coupled to the filter cavity as a reflective component. However, the reflectivity of silicon nitride membranes is typically low ($\approx 0.2$). This fact makes their use as a reflective component in the filter cavity impractical due to the high added optical loss. In this section, we show how a pumped membrane dispersively coupled to the coupled-cavity system (i.e. using the membrane-in-the-middle technique~\cite{Jayich2008}) leads to the phase-insensitive amplification of the signal field. In the analysis, we consider standard equations for field propagation and interaction at an optical component. The quantum amplification occurs when an optical field interacts with the membrane which is driven by the radiation pressure force from the beat of the pump and signal fields. The optical fields are defined in Fig.~\ref{fig:exp}.

Interference on the input test mass in the main cavity is given by the equations
\begin{equation}
\begin{split}
    & a_{\rm 1}(t) = r_0 a_{\rm 2}(t-\tau/2) + t_0 \chi a_{\rm f3}(t-\tau_f/2) \\
    & a_{\rm f}(t) = -r_0 \chi a_{\rm f3}(t-\tau_f/2) + t_0 a_{\rm 2}(t-\tau/2),
\end{split}
\end{equation}
where $\tau$ and $\tau_f$ are the round-trip times in the main and filter cavities, $r_0$ and $t_0$ are the field reflectivity and transmissivity of the input mirror of the main cavity, and $\chi = e^{i\theta}$, where $\theta$ is the relative carrier phase delay across the filter cavity, tuned to $\pi/2$ to achieve the signal recycling. We keep $\chi$ as a parameter in the following analysis to maintain generality and cover the detuned signal recycling case in future studies.

Microscopic motion of the main cavity causes a small fraction of the static field, $A$, in the main cavity to convert to a time-dependent field near the end mirror according to the equation
\begin{equation}
    a_{\rm 2}(t) = a_{\rm 1}(t-\tau/2) - 2i A \frac{\omega_0}{c} x(t), 
\end{equation}
where $x$ is the displacement of the end mirror and $c$ is the speed of light.
The field returning to the membrane from the main cavity is given by the equation
\begin{equation}
    a_{f4}(t) = \chi a_{f}(t-\tau_f/2).
\end{equation}

The pump field in the filter cavity at frequency $\omega_0 + \omega_p$ converts to frequencies around the carrier field at $\omega_0$ according to the equations
\begin{equation}
\label{eq:osc_fields}
\begin{split}
    a_{f2}(t) & = \frac{t_m a_{f4}(t) + r_m t_f a_{\rm in}(t) - 2i A_{p1}e^{i\omega_p t} r_m \frac{\omega_0+\omega_p}{c} y(t)}{1 - r_m r_f} \\
    a_{f1}(t) &= r_f a_{f2}(t) + t_f a_{in}(t) \\
    a_{f3}(t) &= t_m a_{f1}(t) - r_m a_{f4}(t) + 2i A_{p2}e^{i\omega_p t} r_m \frac{\omega_0+\omega_p}{c} y(t),
\end{split}
\end{equation}
where $y(t)$ is the oscillator motion, $r_f$ and $t_f$ are the field reflectivity and transmissivity of the input mirror of the filter cavity, and $r_m$ and $t_m$ are the field reflectivity and transmissivity of the membrane. The equations above imply that the membrane and the input filter mirror form a low-finesse cavity with an eigenmode at $\omega_0$. The oscillator is driven by a thermal force ($F_{\rm th}$) and back-action force from the beat of the pump with the signal fields ($F_{\rm rad}$) as given by the equation
\begin{equation}
    \ddot{y} + \gamma \dot{y} + \omega_m^2 y = \frac{1}{M}(F_{\rm th} + F_{\rm rad}),
\end{equation}
where $M$ is the mass of the oscillator and $\gamma$ is its damping rate. The thermal force adds an unwanted noise to the system as discussed in Sec.~\ref{sec:quantum}. The radiation force helps achieve the quantum amplification and is given by the equation
\begin{equation}
\begin{split}
    F_{\rm rad} = \frac{1}{c} (&A_{f1} a_{\rm f1}^*(t) e^{i\omega_p t} + A_{f1}^* a_{\rm f1}(t) e^{-i\omega_p t} + \\
    &A_{f2} a_{\rm f2}^*(t) e^{i\omega_p t} + A_{f2}^* a_{\rm f2}(t) e^{-i\omega_p t} - \\
    & A_{f3} a_{\rm f3}^*(t) e^{i\omega_p t} - A_{f3}^* a_{\rm f3}(t) e^{-i\omega_p t} - \\
    &A_{f4} a_{\rm f4}^*(t) e^{i\omega_p t} - A_{f4}^* a_{\rm f4}(t) e^{-i\omega_p t}).
\end{split}
\end{equation}

The terms $A_{f1--f4}$ refer to pump fields in the filter on both sides of the membrane (the numbering is consistent with the signal fields $a_{f1--f4}$ shown in Fig.~\ref{fig:exp}) and are related to the input pump field $A_{\rm p,in}$ according to the equations
\begin{equation}
    \begin{split}
    A_{f1} &= \frac{t_f} {1 - r_f r_m + r_f t_m^2 \mu/(1 + r_m \mu)}  A_{\rm p,in}\\
    A_{f2} &= \left(r_m + \frac{t_m^2 \mu}{1 + r_m \mu} \right) A_{f1} \\
    A_{f3} &= \frac{t_m}{1 + r_m \mu} A_{f1} \\
    A_{f4} &= \mu A_{f3},
    \end{split}
\end{equation}
where $\mu = \exp(-i \omega_p \tau_f)$ determines the additional phase accumulation of the pump field relative to the carrier field.

Solving the equations above in the frequency domain leads to the input-output relationship between $a_{\rm f3}$ and $a_{\rm f4}$ of the form given by Eq.~(\ref{eq:quamp}), where the vacuum fields at frequencies around $\omega_0 + 2\omega_p$ play the role of an additional amplifier noise $n$. 
The exact expression for $G$ is quite 
complicated, but it can be well approximated as 
\begin{equation}
    G(\omega)\equiv\frac{\tilde a_{\rm f3}(\omega)}{\tilde a_{\rm f4}(\omega)}
    \approx 1+\frac{2 i\, g^2 \omega_m \tau_f}{\omega^2 -i\gamma \omega-\omega_m^2}\,. 
\end{equation}
In the above formula, $\omega$ is related to the signal 
sideband frequency, $\Omega$, through $\omega=\Omega-\omega_p$.
The constant $g$ quantifies the optomechanical coupling strength between the signal field and the membrane, and it is approximately equal to
\begin{equation}
    g \approx \left(\frac{ r_m t_m^2 P_f \omega_0}{(1-r_f r_m)^2 M c L_f \omega_m}\right)^{1/2}\,. 
\end{equation} 
Here, $P_f$ is the optical power of
the pump field inside the cavity formed by the membrane and the input mirror and dominates over the optical power on the other side of the membrane. 
Guided by the analysis in this section, we present our choice of experimental parameters in the following section.

\section{Experimental parameters}
\label{sec:par}

The quantum amplification is optimal and stable when the Hamiltonian of the full optomechanical system is parity-time symmetric in the single-mode approximation~\cite{Xiang_2020}. The condition is fulfilled when the coupling strength, $g$, equals the coupled-cavity resonant frequency, $\omega_c$, given by the equation~\cite{NeutronStar2019} 
\begin{equation}
    \omega_c = \frac{c}{2}\sqrt{\frac{{\cal T}_0}{L_0 L_f}}\,,
\end{equation}
The equation above highlights the complexity of the table-top demonstration of the quantum phase-insensitive amplification: for meter-scale cavity lengths, $L_0$ and $L_f$, the coupled-cavity resonance is in the tens of kHz range. Therefore, the optomechanical coupling strength, $g$, must be larger for smaller scale experiments than for km-scale ones.

In practice, $g$ is chosen to be slightly smaller 
than $\omega_c$ to maintain a practical stability 
margin, as the system becomes unstable for $g > \omega_c$. We choose
\begin{equation}
g \approx 0.97 \omega_c\,,
\end{equation}
which is used in the subsequent sensitivity analysis, as this is the highest value of $g$ we confirmed to work in numerical simulations. However, $g$ can be optically tuned over the full range of interest (from 0 to above $\omega_c$), which gives us the ability to experimentally explore the margin of stability and the sensitivity enhancement in more detail.

The mechanical eigenmode frequency must satisfy the condition given by the equation
\begin{equation}
    \omega_m \gg \gamma_f.
\end{equation}
Since the filter cavity bandwidth, $\gamma_f$, must be larger than the highest signal frequency ($\gamma_f > \omega_s$), we choose $\gamma_f$ and $\omega_m$ as shown in Table~\ref{table:par}. The relatively low finesse of the filter cavity is similar to that of the Advanced LIGO signal recycling cavity and helps mitigate the negative consequences of the optical losses.

\begin{figure}
\includegraphics[width=\columnwidth]
{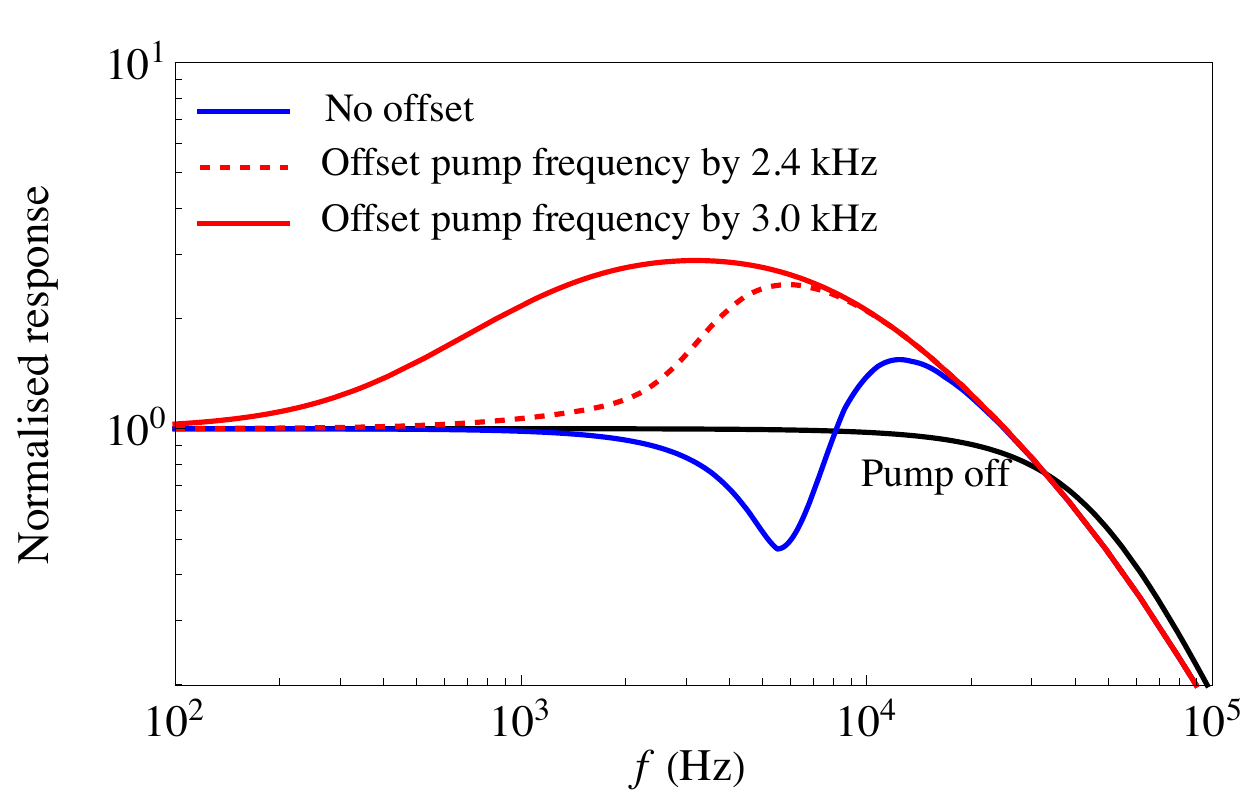}
\caption{Plot showing the effect of offsetting the pump frequency on the signal response. The signal response is maximised when the offset is equal to the optical spring frequency shown in Eq.\,\eqref{eq:os}.}
\label{fig:resp}
\end{figure}

\begin{figure}
\begin{overpic}[width=\columnwidth]
{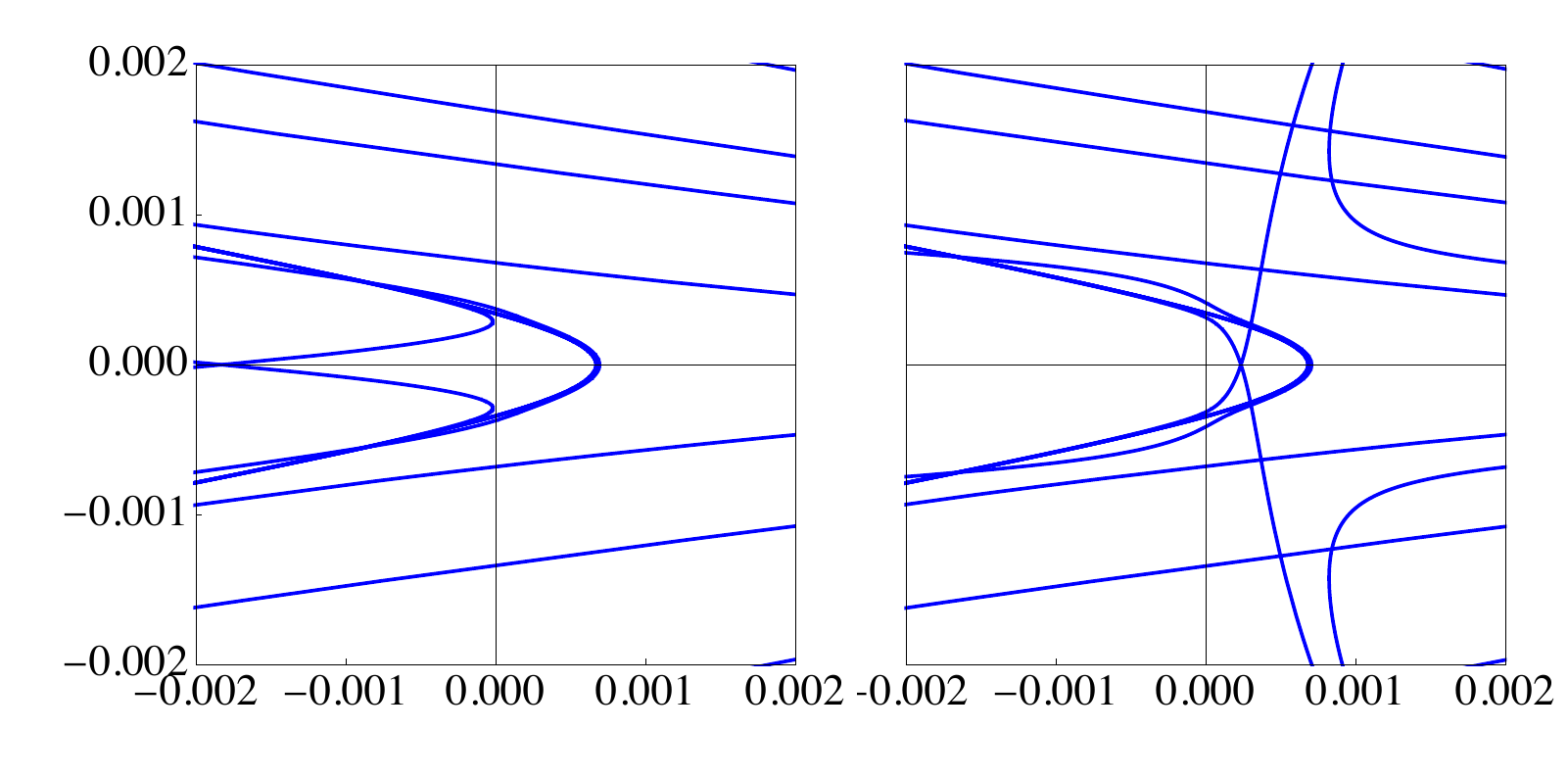}
\put(0,150){\rotatebox{90}{$\Im[\det({\bf I} + {\bf M}_{\rm OL})]$}}
\put(400,5){{$\Re[\det({\bf I} + {\bf M}_{\rm OL})]$}}
\end{overpic}
\caption{Figures showing the Nyquist plots of the system. The left (right) figure shows the case without (with) offsetting the pump frequency. The contour does not enclose the origin after we offset the pump frequency by the optical frequency, which implies a stable system. }
\label{fig:nyq_os}
\end{figure}

The pump field must be tuned to the resonant frequency of the membrane in the presence of the optical fields. The fields stiffen the mechanical oscillator due to the optical spring effect, which results in a shift of the resonant frequency, given by the equation
\begin{equation}\label{eq:os}
 \Delta \omega_{\rm OS}= \frac{r_m t_m^2 g^2 \omega_m}{4(1-r_f r_m)^2(\gamma_f^2 + \omega_m^2)}\,,   
\end{equation}
and is approximately equal to $3.0\,\rm kHz$ for our set of parameters. Such a frequency offset is crucial for both improving the sensitivity and maintaining the stability of the system. In Fig.\,\ref{fig:resp}, we illustrate its effect on the signal response, which is normalized to unity at low frequencies for clarity. In Fig.\,\ref{fig:nyq_os}, we follow Ref.\,\cite{OptomechTD2021} and 
show the Nyquist plot---a parametric plot of the imaginary and real part of the 
determinant of ${\bf I}+{\bf M}_{\rm OL}$ with $\bf I$ being the $2\times 2$ identity matrix and ${\bf M}_{\rm OL}$ the open-loop transfer matrix for the signal field and the idler field. 
Physically, the frequency offset of the pump field, $\omega_p$, is tuned with a pair of acousto-optical modulators (AOMs) as shown in Fig.~\ref{fig:exp}. 

Before measuring the quantum sensitivity, the interferometer must to tuned on resonance at $\omega_0$. We propose to stabilise the filter and the main cavity relative to the probe field (shown in red in Fig.~\ref{fig:exp}) using the Pound-Drever-Hall scheme.
Similar to the Advanced LIGO detectors~\cite{AdvLIGO}, we can stabilise the detector by resonating radio-frequency sidebands in the filter cavity. The resonance is needed to diagonalise the coupled longitudinal degrees of freedom: the filter and main cavities. Our parameter choice implies that the first radio modulation frequency (equal to the free spectral range $c/(2L_f$) of the filter cavity) should be 75\,MHz. For our set of parameters, we get 600\,mW of carrier power in the main cavity and 5\,uW of carrier power in the filter cavity for the input carrier power of 1 mW. The second modulation frequency is set to 10\,MHz to avoid resonances in either cavity. Therefore, we can diagonalise the two degrees of freedom according to the sensing matrix as shown in Table~\ref{table:sensing}.

Stabilisation of the main cavity relative to the pre-stabilised laser can be achieved by demodulating the 10\,MHz sideband and actuating on the laser frequency in a high-gain feedback loop. High bandwidth is required to keep stability of the loop when we introduce the pump field. The 75\,MHz signal is used to stabilise the filter cavity through actuation on the input filter mirror with a bandwidth of 100\,Hz. The pump field leads to the amplification in the filter cavity and increases the optical gain of the detector in the 100\,Hz--20\,kHz band. Therefore, feedback servos are used to maintain stability across both regimes---with and without the pump field.

\begin{table}[t]
\begin{center}
\caption{\label{table:sensing}Optical gains in arbitrary units of the sensors in response to the excitation of the main and filter cavities.}
\begin{ruledtabular}
\begin{tabular}{lcc}
 Demodulation frequency & Main cavity & Filter cavity \\
 \hline
 75\,MHz &	1 &	-1/3 \\
 10\,MHz &	-1 & 0
\end{tabular}
\end{ruledtabular}
\end{center}
\end{table}

\section{Quantum Sensitivity}
\label{sec:quantum}

In this section, we show that improvements in the quantum-limited sensitivity can be achieved with current technology. In the GW detectors, complex seismic isolation systems are installed to suppress ground vibrations, whereas such advanced systems are not present in our design. In the table-top experiment, we thus tune $\omega_c$ and the other experimental parameters discussed in Sec.~\ref{sec:par} to achieve the quantum-limited sensitivity improvements at 100\,Hz--20\,kHz. This is a crucial design choice to avoid coupling of the ground vibrations to the experiment. As a result, the setup is limited by the thermal noise of the membrane and vacuum fields from the interferometer's open ports: the input port and loss channels in the filter and main cavities.

The effect of the amplification can be classically demonstrated by measuring the transfer function from the main cavity end mirror motion to the readout photodetectors. However, the amplification of the optical gain can be canceled by the noise amplification if the optical and mechanical parameters are not chosen carefully. Our proposal to avoid this excess noise amplification uses a high-quality-factor Si$_3$N$_4$ membrane operated at a cryogenic temperature, which has been recently used in 
quantum-limited measurements~\cite{Rossi2018, Mason2019}. 

\begin{figure}[b]
\includegraphics[width=\columnwidth]{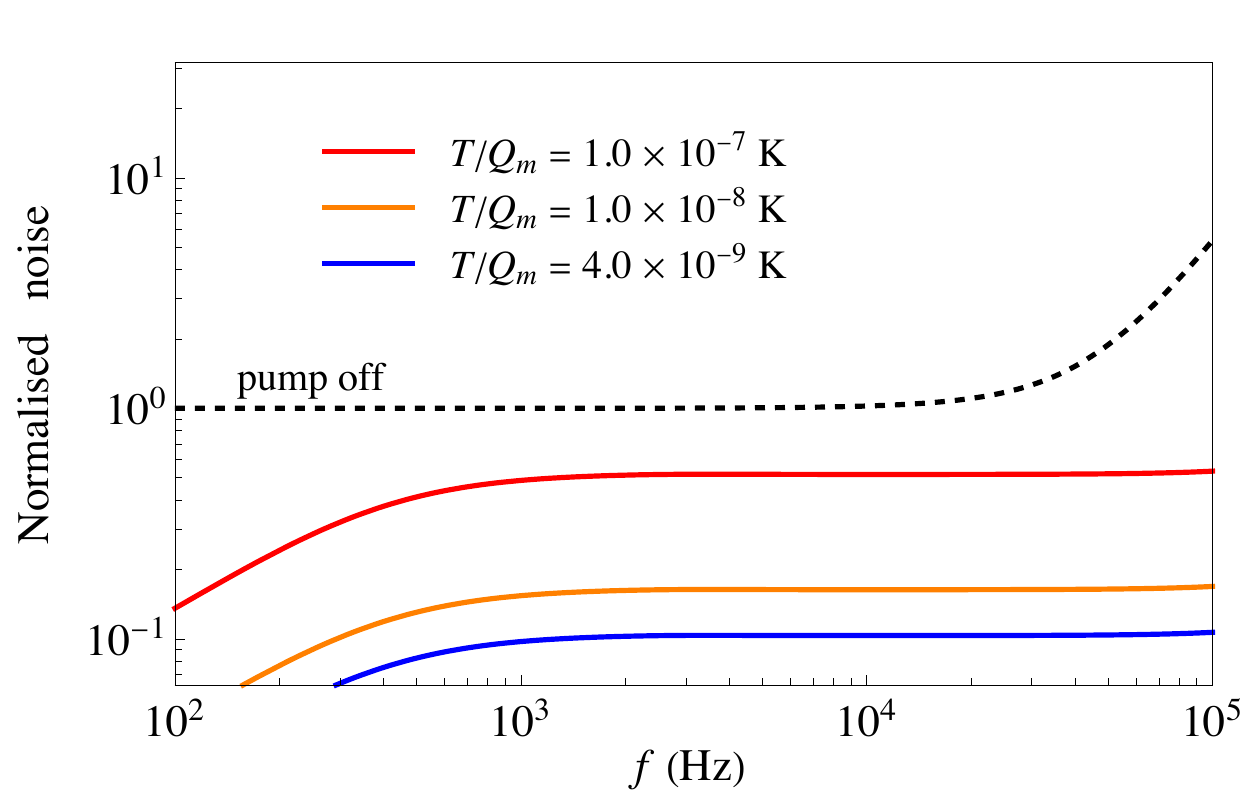}
\caption{Figure showing the thermal noise contribution of the membrane at different environmental temperature. As a reference, the dotted curve shows the 
quantum noise level when the pump power on the membrane is turned off. All the curves are normalised with respect to the low frequency (below 100\,Hz) noise level in the pump-off case. } 
\label{fig:thermal_noise}
\end{figure}

Fig.~\ref{fig:thermal_noise} shows the sensitivity of the same system under consideration for different levels of thermal noise, specified by the ratio $T / Q_m$. Given this result, it is possible to see real sensitivity improvements for a $T/Q_m$ ratio as large as $1.0\times 10^{-7}\,\rm K$. A significant improvement in the sensitivity can be achieved for $T/Q_m < 1.0 \times 10^{-8}$. With a reasonable temperature of 10\,K, this implies a constraint of $Q_m\ge 1.0\times 10^9$. This parameter regime is achievable with the state-of-the-art Si$_3$N$_4$ membranes.

Optical losses in the main cavity come from scattering and absorption of the laser beam on the optical coating. For our metre-scale setup, optical losses as small as $5$\,ppm per mirror are routinely achieved by commercially available superpolished mirrors~\cite{Realistic}. We follow the formalism discussed in~\cite{Miao2018} and find the noise level imposed by the loss $\epsilon_0$ in the main cavity:
\begin{equation}
S_0^{\epsilon}(\omega) = \frac{4}{{\cal T}_{\rm eff}} \epsilon_0\,,
\end{equation}
where the noise's power spectral density (PSD) $S_0^{\epsilon}$ is normalized to
the DC level of the noise PSD in the absence of the pump field (i.e. the one provided by the shot noise only), and ${\cal T}_{\rm eff}$ is the effective power transmissivity of the compound mirror formed by the central mirror, the membrane, and the input mirror. 

The filter cavity, similar to the recycling cavities in Advanced LIGO, witnesses larger optical losses due to a larger number of mirrors, the anti-reflective coating of the main cavity input coupler, and mode mismatch between the two cavities. However, the filter cavity loss, $\epsilon_f$, is also less important at low frequencies. If we keep the lowest order to the filter cavity loss, its contribution to the noise level is approximately given by:
\begin{equation}
S_f^{\epsilon}
(\omega)=\frac{2(\gamma_0^2+\omega^2)\tau}{{\cal T}_{\rm eff}\,\gamma_0}\epsilon_f\,, 
\end{equation}
where $\gamma_0\equiv {\cal T}_0/(2\tau)$ is the bandwidth of the main cavity. 

\begin{figure}[h]
\includegraphics[width=\columnwidth]{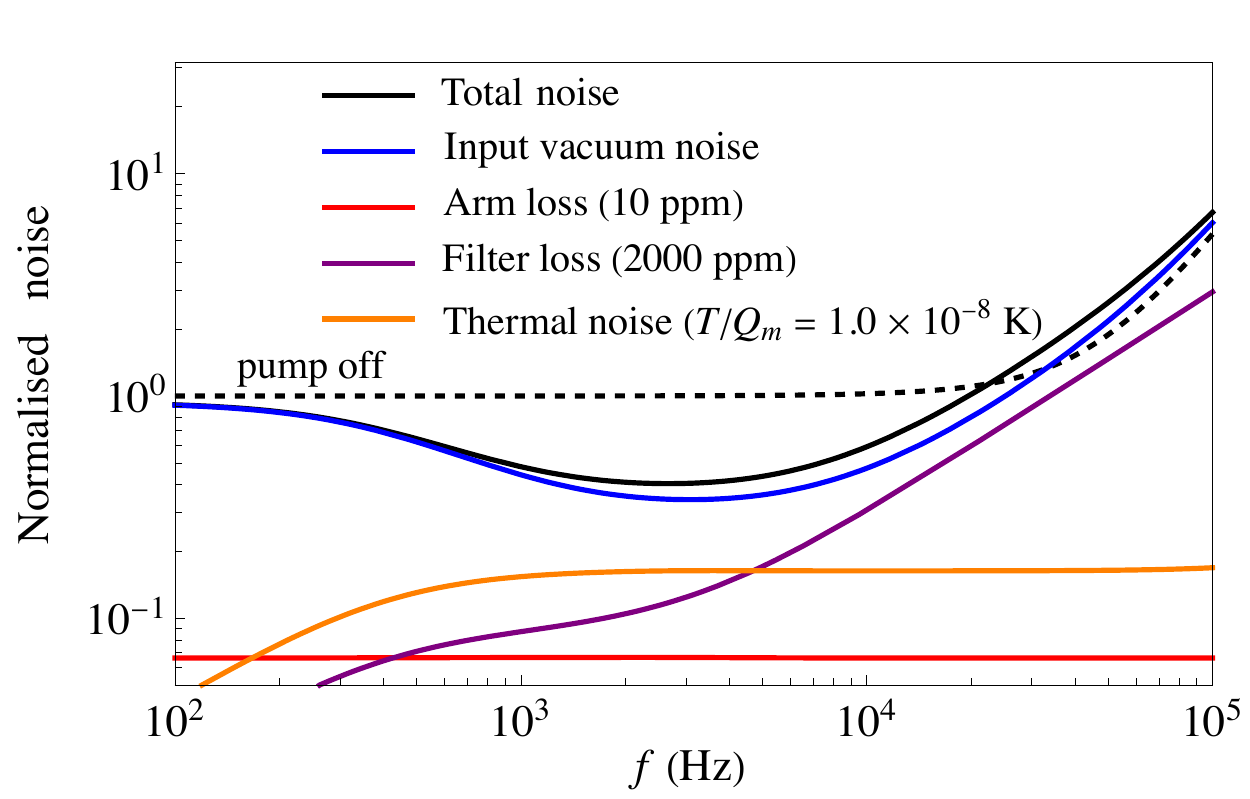}
\caption{Figure showing the noise budget of our proposed experiment for the quantum demonstration with realistic optical loss both in the main cavity and the filter cavity. The thermal noise from the membrane is also included. The ``input vacuum noise" arises from  the vacuum fluctuation at the input; it is the lowest quantum noise level when the optical loss and thermal noise are absent.} 
\label{fig:total_noise}
\end{figure}

Optical-loss-induced dissipation is analogous to the sensitivity limit from the mechanical dissipation. The thermal noise can be mapped to an equivalent optical loss in the main cavity by using the following equation\,\cite{Miao2018}: 
\begin{equation}
    \epsilon^{\rm eq}_{0} = \frac{k_B \gamma_0}{\hbar g^2}\left(\frac{T}{Q_m}\right)\,. 
\end{equation}
With the parameters listed in Table\,\ref{table:par}, $T/Q_m \sim 10^{-8}\,\rm K$ corresponds to around 70 ppm loss in the main cavity.

Fig.\,\ref{fig:thermal_noise} shows 
the total noise of our proposed experiment. It includes additional 
quantum noise from the optical loss in 
the main cavity and filter cavity. As we can see, given 
a realistic level of optical loss and $T/Q_m$ ratio of $10^{-8}\,{\rm K}$, we could observe a factor of two suppression in the noise level compared to the pump-off case, which is comparable to the quantum noise improvement from squeezed states of light.

The state-of-the-art membranes can achieve even lower values than we need, $T/Q_m \leq 10^{-10}$ K, if positioned in a dilution refrigerator~\cite{Yuan_2015}. However, the optical absorption and low thermal conductivity of silicon nitride will increase the membrane temperature~\cite{Leivo_1998, Ftouni_2015} above mK temperatures. We estimate the thermal resistance of a membrane heated by an optical beam at its center as
\begin{equation}
\label{eq:R}
    R \approx \frac{R_m}{w} \frac{1}{2 \pi h}
    \frac{1}{\alpha(T)},
\end{equation}
where $R_m / w = 2.5$ is the ratio of the membrane radius to the beam size on the membrane, $\alpha(T)$ is the temperature-dependent coefficient of thermal conductivity, and $h$ is the membrane's thickness. The membrane temperature is then given by the equation
\begin{equation}
\label{eq:T}
    T \approx R P_a,
\end{equation}
where $P_a$ is the power absorbed by the membrane. We compute the power according to the equation
\begin{equation}
    P_a = A  B  P_f,
\end{equation}
where $A = 10$\,ppm is the membrane absorption computed using the imaginary part of the film refractive index, $k = 10^{-5}$~\cite{Pan_2018}, and $B \approx 6 \times 10^{-3}$ is the ratio of the power on the membrane to the maximum filter power. Similar to~\cite{Thompson_2008}, we propose to minimise $B$ by positioning the membrane at the node of the cavity field.

Solving Eq.~\ref{eq:R} and~\ref{eq:T} relative to the membrane temperature $T$ and approximating $\alpha = 0.23 + 0.032 (T-T_0)$\,W/(mK) around $T_0=10$\,K~\cite{Ftouni_2015}, we get the minimum possible membrane temperature of $T = 8.6$\,K with the parameters listed above and in Table.~\ref{table:par}. Therefore, we can assume the membrane temperature of $T=10$\,K in our estimations of the thermal noise shown in Fig.~\ref{fig:thermal_noise}.

\section{Conclusion}
\label{sec:conc}

Active parity-time-symmetric and phase-insensitive quantum amplification stands as a promising method of enhancing the quantum-limited sensitivity of interferometric devices. We propose an experiment that has the potential to demonstrate quantum amplification using existing technology on a tabletop scale. This is a crucial step towards embedding the technology in devices such as gravitational-wave and axion detectors. The technology has the potential to increase the reach of the detectors by an order of magnitude~\cite{OptomechTD2021}.

We optimise the experiment to demonstrate quantum amplification with a pair of coupled high-finesse and low-finesse optical cavities. The quantum amplification of the signal field can be achieved with an Si$_3$N$_4$ membrane at $10\,$K and a properly tuned pump beam. Our design targets the suppression of quantum noises around 100\,Hz--20\,kHz to avoid noises related to ground and acoustic vibrations.

We explore the stability of such a system in terms of the optomechanical coupling strength, the pump-field frequency, and maintenance of the cavities' operating point. To this end, we lay out the methods for control of the pump-field frequency using a dual AOM approach, and for the cavity length control using two radio-frequency sidebands in a Pound-Drever-Hall locking scheme. We thus specify a practical solution for the stable control of such a system. Through the setting of the pump-field amplitude (optical control), we are have the ability to alter the amplifier gain in real-time without disrupting operation.

We show that quantum noises from realistic optical losses in the main and filter cavities do not negate the positive effect of the quantum amplification. The choice of the mechanical oscillator (Si$_3$N$_4$ membrane) lets us avoid significant clipping losses in the filter cavity. Moreover, Si$_3$N$_4$ membranes have applications beyond tabletop experiments in km-scale gravitational-wave detectors.

\section{Acknowledgements}
We thank members of the LIGO QNWG group for useful discussions and Joe Bentley for his valuable feedback. J.S., A.D, and D. M. acknowledge the support of the Institute for Gravitational Wave Astronomy at the University of Birmingham, STFC Quantum Technology for Fundamental Physics scheme (Grant No. ST/T006609/1), and EPSRC New Horizon Scheme (Grant No. EP/V048872/1). H. M. is supported by State Key Laboratory of Low Dimensional Quantum Physics and Frontier Science Center for Quantum Information at Tsinghua. D.M. is supported by the 2021 Philip Leverhulme Prize. C.Z. acknowledge the support of the Australian Research Council (grant No. DP170104424 and CE170100004).

\bibliography{qamp}

\end{document}